\documentclass[12pt]{article}

\usepackage[super,comma,sort&compress]{natbib}
\usepackage{amsmath}
\usepackage{amssymb}
\usepackage{color}
\usepackage{graphicx}
\usepackage{multirow}
\usepackage{url}
\usepackage[margin=1in]{geometry}
\usepackage[usenames,dvipsnames]{xcolor}
\usepackage{fancyhdr}
\usepackage{setspace}
\begin{document}

\title{Dissecting the roles of local packing density and longer-range effects in protein sequence evolution}

\author{Amir Shahmoradi$^{1,2,3}$\footnote{Correspondence to: amir@physics.utexas.edu}~, Claus O. Wilke$^{2,3,4}$\footnote{Correspondence to: wilke@austin.utexas.edu}}
\date{}
\maketitle

\noindent
$^1$ Department of Physics, The University of Texas at Austin\\
$^2$ Center for Computational Biology and Bioinformatics, The University of Texas at Austin\\
$^3$ Institute for Cellular and Molecular Biology, The University of Texas at Austin\\
$^4$ Department of Integrative Biology, The University of Texas at Austin\\

\begin{abstract}
What are the structural determinants of protein sequence evolution? A number of site-specific structural characteristics have been proposed, most of which are broadly related to either the density of contacts or the solvent accessibility of individual residues. Most importantly, there has been disagreement in the literature over the relative importance of solvent accessibility and local packing density for explaining site-specific sequence variability in proteins. We show that this discussion has been confounded by the definition of local packing density. The most commonly used measures of local packing, such as contact number and the weighted contact number, represent the combined effects of local packing density and longer-range effects. As an alternative, we propose a truly local measure of packing density around a single residue, based on the Voronoi cell volume. We show that the Voronoi cell volume, when calculated relative to the geometric center of amino-acid side chains, behaves nearly identically to the relative solvent accessibility, and each individually can explain, on average, approximately 34\% of the site-specific variation in evolutionary rate in a data set of 209 enzymes. An additional 10\% of variation can be explained by non-local effects that are captured in the weighted contact number. Consequently, evolutionary variation at a site is determined by the combined effects of the immediate amino-acid neighbors of that site and effects mediated by more distant amino acids. We conclude that instead of contrasting solvent accessibility and local packing density, future research should emphasize on the relative importance of immediate contacts and longer-range effects on evolutionary variation.
\end{abstract}


\section*{Introduction}

A variety of site-specific structural characteristics have been proposed over the past decade to predict protein sequence evolution from structural properties. Among the most important and widely discussed are the Relative Solvent Accessibility (RSA) \cite{Overingtonetal1992, goldman_assessing_1998, MirnyShakhnovich1999, bustamante_solvent_2000, conant_solvent_2009, franzosa_structural_2009, ramsey_relationship_2011, scherrer_modeling_2012, meyer_integrating_2013, meyer_cross_species_2013, yeh_site-specific_2014, yeh_local_2014, shahmoradi_predicting_2014, sikosek_biophysics_2014, meyer_geometric_2015}, Contact Number (CN) \cite{rodionov_sequence_1998, hamelryck_amino_2005, liao_protein_2005, bloom_structural_2006, huang_mechanistic_2014, marcos_too_2015, yeh_site-specific_2014, yeh_local_2014, shahmoradi_predicting_2014}, measures of thermodynamic stability changes due to mutations at individual sites in proteins \cite{MirnyShakhnovich1999, DokholyanShakhnovich2001, Liberlesetal2012, echave_relationship_2014, Dasmehetal2014}, protein designability \cite{england_structural_2003, Jacksonetal2013}, and measures of local flexibility, such as the B factor \cite{liao_protein_2005, shih_evolutionary_2012, shahmoradi_predicting_2014}, or flexibility measures based on elastic network models \cite{liu_sequence_2012, huang_mechanistic_2014} and Molecular Dynamics (MD) simulations \cite{shahmoradi_predicting_2014}, or direct comparison of fluctuations among static structures of homologous proteins \cite{fuglebakk_measuring_2012,shahmoradi_predicting_2014, carvalho_comparison_2015}. (For a recent review, see Ref.~\cite{echave_causes_2016}.) It is conceivable that the majority of these site-specific quantities predicting evolutionary rate (ER) simply act as different proxy measures of an underlying biophysical constraint, such as the strength of amino acid interactions at individual sites. Indeed, Huang et al. \cite{huang_mechanistic_2014} recently proposed that the underlying constraint is mechanistic stress, which can be estimated by calculating the weighted contact number (WCN) at each site. Quantities such as relative solvent accessibility and local flexibility are expected to correlate with WCN, because all these measures depend to some extent on the density of contacts around an amino acid\cite{echave_causes_2016}. Huang and coworkers thus argued that the \emph{local packing density} is the primary constraint of protein sequence evolution \cite{huang_mechanistic_2014, yeh_site-specific_2014, yeh_local_2014}.

However, a priori it is not clear whether quantities based on the contact-number concept (either CN or WCN) are truly measures of \emph{local} packing density. All these measures involve adjustable free parameters in their definitions, and depending on the parameter choices, the measures can incorporate substantial long-range effects. In particular, WCN by default uses an inverse-square weighting, which assumes that amino acids at an arbitrary distance away from the focal residue still exert some measurable effect on that residue. Thus, some of the best predictors of evolutionary variation seem to confound local and longer-range effects, and the relative importance of the immediate neighborhood of an amino acid vs.\ longer range effects is not known. Importantly, by longer-range effects we mean here effects between amino acids that are distant in 3D space. Amino acids that are distant in sequence but close in 3D space, as are frequently observed\cite{plaxco_contact_1998,lai_sequence_2015}, would contribute to local packing but not to longer-range effects by our definition.

To separate out the effects of direct-neighbor and longer-range effects, we here derive a new set of structural characteristics that, unlike CN and WCN, depend only on the immediate contacts of a given amino acid and do not involve freely adjustable parameters in their definitions. We achieve this goal by employing tessellation methods developed in the field of computational geometry. We find that quantities based on tessellation methods explain much but not all of the variance in ER that WCN captures. Therefore, we conclude that longer-range effects beyond the first coordination shell play a significant role in protein sequence evolution.

\section*{Methods}

\subsection*{Protein data set and site-specific sequence variability}

We analyzed the same data set of 209 monomeric enzymes we have previously analyzed \cite{echave_relationship_2014} and that was originally published with four additional proteins \cite{yeh_site-specific_2014}. The enzymes were originally randomly picked from the Catalytic Site Atlas $2.2.11$ \cite{porter_catalytic_2004}, range from 95 to 1287 amino acids in length, and include representatives from all six main EC functional classes \cite{webb_enzyme_1992} and domains of all main SCOP structural classes \cite{murzin_scop:_1995}. Evolutionary rates at all sites in all proteins were calculated as described \cite{echave_relationship_2014}. In brief, we employed the software Rate4Site \cite{mayrose_comparison_2004}, using the empirical Bayesian method and the amino-acid Jukes-Cantor-like (JC-like) mutational model.

\subsection*{Packing density and relative solvent accessibility}

Local Packing Density (LPD) for individual sites in all proteins was calculated according to the two most-commonly used definitions, Contact Number (CN) and Weighted Contact Number (WCN) (defined in Eqs.~\ref{eqn:cn} and \ref{eqn:wcn_pwrl} in Results). Both CN and WCN generally require each residue to be represented by a single point in space
(but see the work by Marcos and Echave \cite{marcos_too_2015}, who used a two-point representation). Here, we examined seven different ways to choose the reference point for each amino acid: We considered all backbone atomic coordinates, i.e., the coordinates of N, C, O, or C{$_\alpha$}, the coordinates of the first heavy atom in the side-chains, C{$_\beta$}, and two more representations obtained by averaging over the coordinates of all heavy atoms in the amino acid including the backbone atoms (denoted as AA coordinates) or of all side-chain heavy atoms (denoted as SC coordinates). For each of these seven representations, we calculated CN and WCN for each residue in each protein. Appendix~A shows that the SC representation results in the highest correlations between structural measures and sequence evolution, and therefore we used SC coordinates for all results reported in the main body of the manuscript.

We used the DSSP software \cite{kabsch_dictionary_1983} to calculate  the Accessible Surface Area (ASA) for each residue in each protein, and we normalized these ASA values by their theoretical maximum \cite{tien_maximum_2013} to obtain Relative Solvent Accessibility (RSA). Throughout this work, we considered RSA rather than ASA, consistent with common practice in the literature. RSA is preferred in evolutionary analyses because ASA confounds residue size and residue exposure. The maximum ASA varies by a factor of three from the smallest amino acid (glycine) to the largest (tryptophan) \cite{tien_maximum_2013}. As a consequence, a fully exposed glycine may have the same ASA as a largely but not completely buried tryptophan. By contrast, an RSA of 1 indicates that the residue is fully exposed, regardless of the nature of the residue. Thus, we generally expect RSA to be less variable under amino-acid substitution than ASA. Fully exposed residues tend to remain fully exposed, partially exposed residues tend to remain partially exposed, and so on.

\subsection*{Voronoi tessellation}

We used the VORO++ software \cite{rycroft_voro++:_2009} to calculate the relevant Voronoi cell properties of all sites in all proteins in the dataset. Specifically, we calculated the length of the cell edges, surface area and volume, number of faces of each cell, and the cell eccentricity, defined as the distance between the cell's seed and the geometrical center of the cell. We also calculated the cell {\it sphericity}, which is a measure of the compactness of the cell, defined as
    \begin{equation}
        \label{eqn:sphericity}
        \Psi = \frac{\pi^{\frac{1}{3}}(6V)^{\frac{2}{3}}}{A}.
    \end{equation}
Here, $V$ and $A$ represent the volume and the area of the cell, respectively.  By definition, $\Psi$ falls between 0 and 1. The limit of $\Psi=1$ indicates that the cell is perfectly spherical, while the limit of $\Psi=0$ indicates that the cell is a 2-dimensional object that has no volume but only surface area.

Each Voronoi cell is a polyhedron consisting of a set of polygons joined together at their edges. The total edge length of a cell was calculated as the sum of the edge lengths of all such polygons. The resulting sum was divided by two to obtain the total edge length, since each edge is shared between two polygons. Similarly, the cell surface area was calculated as the sum of the surface areas of all polygons that bounded it. Finally, the cell volume was defined as the volume of the Voronoi polyhedron.

As for CN and WCN, we considered seven different coordinate representations, and we found that the SC coordinates performed best (Appendix A). Therefore, unless otherwise noted, all results reported in the main body of the text are based on SC coordinates.

\subsection*{Edge effects in Voronoi partitioning of protein structures}

When performing a Voronoi tessellation of a protein, which has a finite size, we have to be aware of the existence of {\it edge effects} on the protein surface. Most Voronoi cells on the protein surface are not completely enclosed by other neighboring cells. Therefore, to ensure that all cells are closed in all directions, we need to place the protein structure into a hypothetical box of arbitrary size. This box provides the missing boundaries for the otherwise open cells on the protein surface. Voronoi cells that are bounded by this box tend to have artifactual properties, such as much larger volumes and areas compared to the cells that are fully surrounded by other neighboring cells. In addition, the volumes and areas of these cells depend on the size of the box, which is arbitrary.

To verify that these edge effects did not influence the observed sequence-structure correlations, we investigated how our results changed as a function of the size of the enclosing box. To do so, we first identified all open cells in all proteins by examining the fraction of change in the cell volumes upon changing the size of the containing box.  Initially, each wall of the box was set to minimum distance of $30.0$\AA\ from all atoms in the protein, and then the size of the box was changed, such that each wall would be at least $60.0$\AA\ away from any atom in the protein.   Note that changing the size of the box surrounding the protein will only affect the properties of those cells that are inherently open in the absence of the box.

For each protein, once open cells were identified, we sorted these cells according to their rates of volume change upon changing the box size, such that the slowest changing cell was assumed to have the smallest volume among open cells, and the fastest changing cell was assumed to have the largest volume among open cells. Note that this method of ranking the open cells is very similar to ranking the cells by the solid opening angle of open cells, without the requirement of explicitly measuring the solid opening angles for all open cells. The underlying assumption here is that cells with larger solid opening angles will show larger rate of cell volume change upon changing the box size surrounding the protein. The wider the solid opening angle of a cell is, the more space is assumed to be available to the amino acid residing the open cell. In essence, the rate of volume change provides a method of ranking the open cells based on their solid opening angles, without direct calculation of the angles.

All open cells were assumed to have a larger volume than any closed cell.  Finally, we calculated Spearman's correlation coefficient between this ordered set of cell volumes (that included both open and closed cells) and the evolutionary rates. We then compared the computed correlation coefficients to those obtained by using the original cell volumes without corrections for edge effects. The resulting distributions of correlation coefficients for the original and corrected cell volumes were nearly identical (mean difference $<0.001$).

For our dataset of 209 monomeric proteins, we thus concluded that edge effects due to Voronoi tessellation appear to have virtually no influence on the observed sequence-structure correlations. We reached similar conclusions when open cells were alternatively ranked by other criteria, such as the fractional changes in cell area upon changing the box size. Thus, the Voronoi cell characteristics, in particular cell volume and cell area, can be safely used in predicting sequence variability, as long as a sufficiently large enclosing box is used to calculate the Voronoi cell characteristics of open cells.

One exception, however, is cell sphericity, as defined in Eqn. \ref{eqn:sphericity}. This quantity behaves differently for open and closed cells. Unlike the case for closed cells in the protein's interior,  the sphericity of open cells on the surface of the protein is generally {\it positively} correlated with site-specific evolutionary rates. This is an artifact of the edge effects in the tessellation of finite structure of protein. A possible correction to the sphericity of open cells could be therefore applied by negating the value of sphericity for open cells. This correction resulted on average in a 0.05 increase in the correlation coefficients of sphericity with ER. However, since this correction has no obvious justification based on first principles, we did not employ it for further analyses. More importantly, the inclusion or exclusion of edge-effect corrections to cell sphericity did not substantively affect any of the major findings of this work.

\section*{Statistical analysis and data availability}

Statistical analyses were carried out with the software package R \cite{ihaka1998r}. We used Spearman correlations and partial Spearman correlations throughout this work. Partial Spearman correlations were calculated using the R package {\it ppcor} \cite{kim_understanding_2006}.

All data and analysis scripts necessary to reproduce this work are publicly available to view and download at \url{https://github.com/shahmoradi/cordiv}.

\section*{Results}

\begin{figure}[h]
        \begin{center}
            \includegraphics[width=6in]{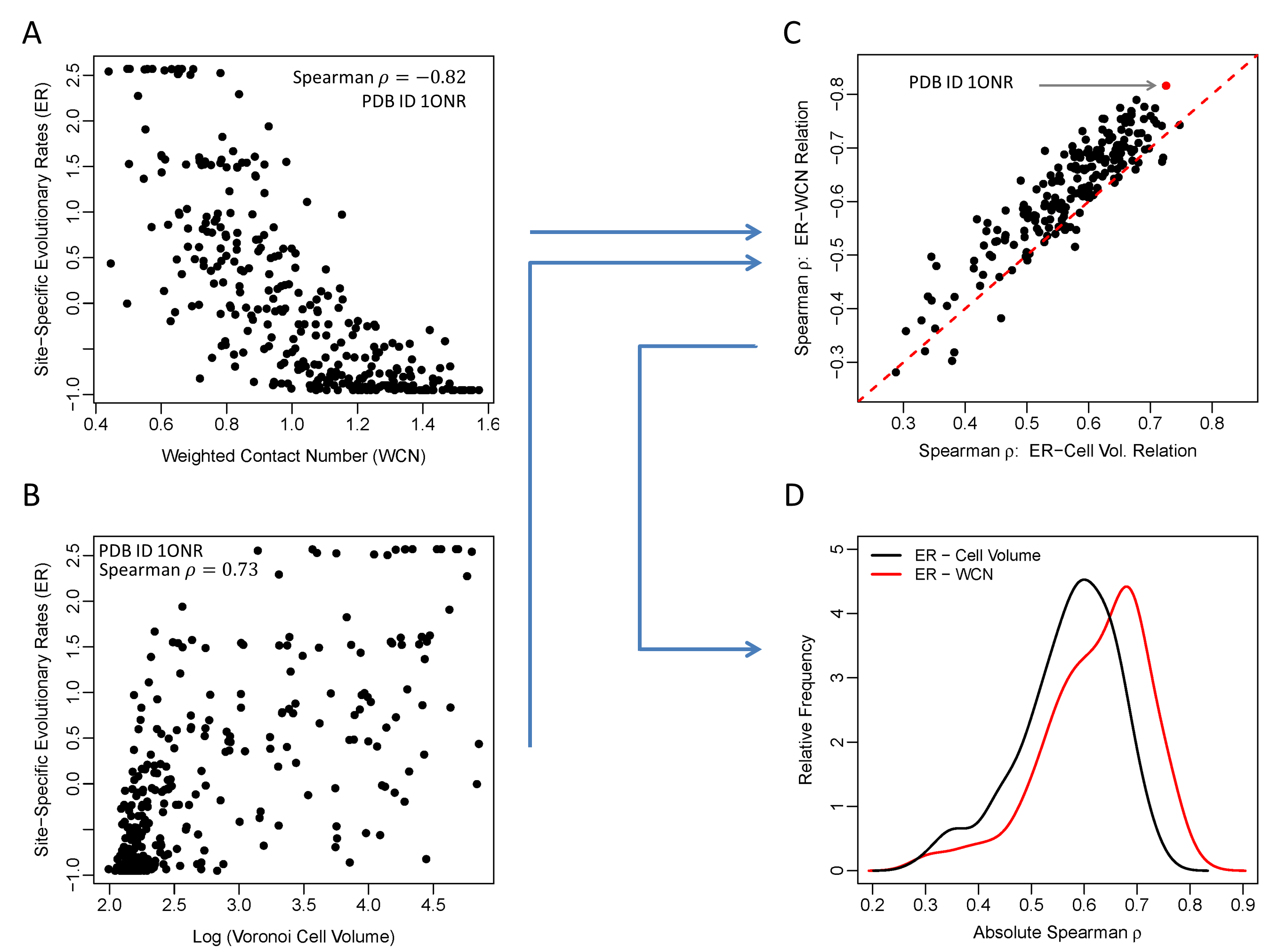}
        \end{center}
        \caption{Schematic of our analysis approach. (A, B) Evolutionary rates at each site in a protein correlate with measures of packing density, such as WCN (A) or Voronoi cell volume (B). Results are shown here for all amino acids in one representative protein (PDB ID: 1ONR, chain A) (C) For all proteins in the data set, the Spearman's correlation coefficients of ER--WCN and ER--Cell Volume relations are calculated and compared to each other. Each black point represents the two correlation values for one protein. The red dashed line represents the equality line for the absolute values of the correlation strengths. (D) Finally, we convert the set of correlation coefficients into distributions and compare their relative means. On average, WCN correlates more strongly with ER than cell volume does.}
        \label{fig:pipeline}
    \end{figure}

Throughout this study, we ask to what extent various measures of local packing in proteins predict evolutionary variation. To address this question, we carry out a number of analyses that all consist of the same conceptual steps illustrated in Figure \ref{fig:pipeline}. We first correlate the evolutionary rate (ER) at each site with a measure of packing at the same site (Figure \ref{fig:pipeline}A, B). We then calculate the corresponding correlation coefficients separately for all proteins in our data set (Figure \ref{fig:pipeline}C). Finally, we compare the distributions of these correlation coefficients for different measures of packing (Figure \ref{fig:pipeline}D).

\subsection*{Optimal definition of contact number incorporates longer-range effects}

Prior work on the same enzyme data set that we have analyzed here has shown that contact number generally outperforms all other structural predictors of site variability \cite{yeh_site-specific_2014, yeh_local_2014, huang_mechanistic_2014}. This finding has been interpreted as indicating that local packing density is the primary evolutionary constraint imposed by protein structure. However, to what extent the contact number (CN) is actually a truly local measure is unclear. The contact number at a site $i$, CN$_i$, is defined as the number of amino acids within a fixed radius $r_0$ from that site,
    \begin{equation}
        \label{eqn:cn}
        \text{CN}_{i} = \sum_{j\neq i} H\big(r_0-r_{ij}\big),
    \end{equation}
where $r_{ij}$ is the distance between sites $i$ and $j$ and $H(r)$ is the Heaviside step function
    \begin{equation}
        \label{eqn:heaviside}
        H(r) = \begin{cases}
        			1 \quad\text{for $r\geq 0$,}\\
                    0 \quad\text{otherwise.}
               \end{cases}
    \end{equation}
The location of individual sites is commonly chosen as the position of the residues' C$_\alpha$ backbone atoms, but recent work has shown that the geometric center of the side chain may be a better choice \cite{marcos_too_2015}. Irrespective of the particular choice of reference point for each site, how local of a measure CN$_i$ actually is depends on the arbitrary cutoff parameter $r_0$. There is no consensus on the optimal value of this cutoff distance, although it is typically chosen in the range of 5\AA\ to 18\AA\ \cite{lin_deriving_2008, franzosa_structural_2009, weng_molecular_2014, yeh_local_2014}. While a cutoff of 5\AA\ likely captures only immediate neighbors of the focal amino acid, a cutoff of 18\AA\ will capture many amino acids that do not directly interact with the focal amino acid, and thus captures both local and non-local packing-density information.

\begin{figure}
        \begin{center}
        \begin{tabular}{cc}
            \includegraphics[width=3.1in]{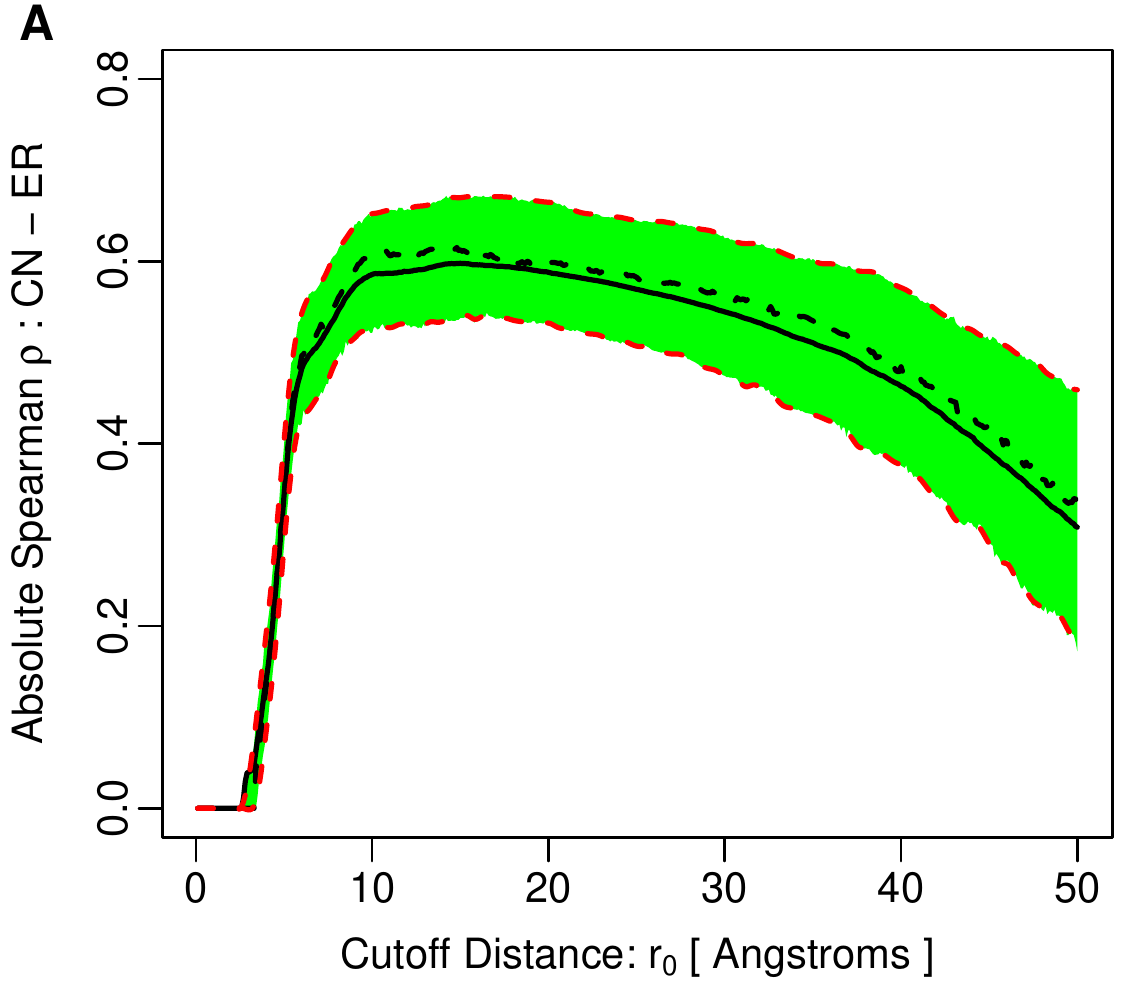} & \includegraphics[width=3.1in]{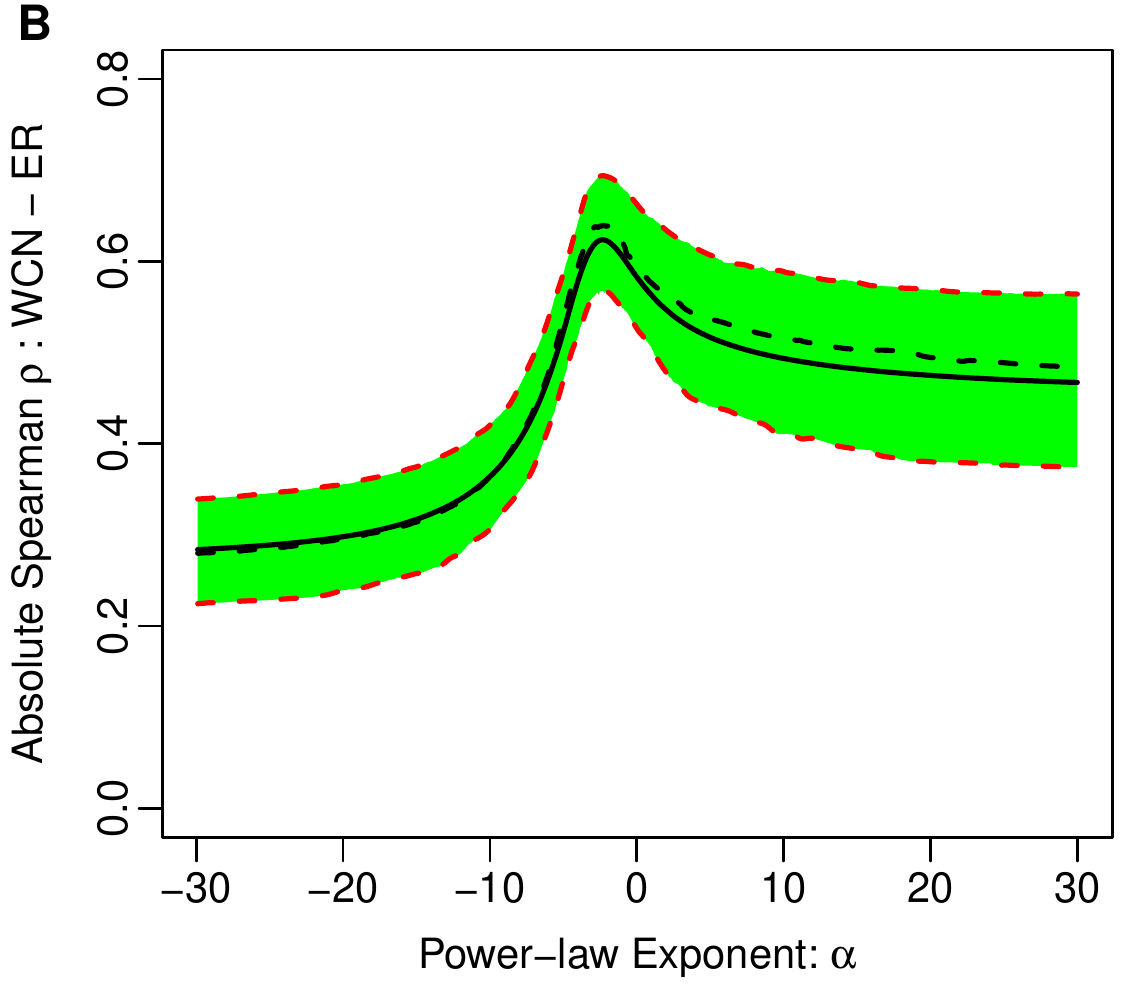}
        \end{tabular}
        \end{center}
        \caption{Absolute correlation of evolutionary rate with (A) Contact Number and (B) Weighted Contact Number for varying degrees of locality of these quantities. In (A), we vary the cutoff parameter $r_0$ in Eq.~\ref{eqn:cn} from 0 to 50\AA. In (B), we vary the exponent $\alpha$ in Eq.~\ref{eqn:wcn_pwrl} from $-30$ to 30. In each plot, the solid black line represents the mean correlation strength in the entire dataset of 209 proteins and the dashed black line indicates the median of the distribution. The green-shaded region together with the red-dashed lines represent the $25\%$ and $75\%$ quartiles of the correlation-strength distribution. Note that for the case of WCN with $\alpha>0$ the sign of the correlation strength $\rho$ is the opposite of the sign of $\rho$ with $\alpha<0$. In addition, $\rho$ is undefined at $\alpha=0$ and not shown in this plot. The parameter values at which the correlation coefficient reaches the maximum over the entire dataset are given in Table \ref{tab:best_params}.}
        \label{fig:cnwcnp}
    \end{figure}

To determine at what cutoff parameter $r_0$ the contact number had the most predictive power for evolutionary variation, we calculated, for each protein in our data set, the Spearman correlation between evolutionary rate (ER) and contact number CN across the protein as a function of $r_0$ (Figure \ref{fig:cnwcnp}A). (A similar analysis has previously been published\cite{yeh_local_2014}.) We found that, on average, the correlation started to rise rapidly around 5\AA, reached a maximum around $14.3$\AA, and then started to decline slowly for larger $r_0$ (see also Table~\ref{tab:best_params}). Since the maximum correlation was located at a cutoff value that captured at least two complete shells of amino acids around the focal residue, this analysis suggested that non-local effects play some role in constraining amino-acid variability.

    \begin{table}[htbp]
    \caption{The first, second (median), and the third quartiles of the distribution of the best free parameters of the Contact Number (CN) and the Weighted Contact Number (WCN) resulting in the strongest correlations with sequence evolutionary rates. The best parameter distributions are obtained by tuning the free parameters of CN and WCN for individual proteins in the dataset of 209 monomeric enzymes such that the strongest correlation with evolutionary rates is obtained. \label{tab:best_params}}
    \bigskip
    \centerline{
    \begin{tabular}{lccc}
      \hline
      \multirow{2}{*}{Local Packing Density Measure} & \multicolumn{3}{c}{Best free-parameter} \\ \cline{2-4}
                                           & $25\%$ quartile & median & $75\%$ quartile  \\ \hline
      CN    & $r_0=9.3~[\text{\AA}]$ & $r_0=14.3~[\text{\AA}]$ & $r_0=19.6~[\text{\AA}]$  \\
      WCN & $\alpha=-2.7$ & $\alpha=-2.3$ & $\alpha=-1.5$  \\
      \hline
    \end{tabular}}
    \end{table}

The arbitrariness of the cutoff $r_0$ in the CN definition has been long recognized, and therefore some authors \cite{lin_deriving_2008} have suggested an alternative quantity known as the Weighted Contact Number (WCN). For a given site $i$ in a protein of length $N$, WCN$_i$ is defined as the sum of the inverse-squared  distances to all other sites in protein,
    \begin{equation}
        \label{eqn:wcn_pwrl}
        \text{WCN}_{i} = \sum^N_{j\neq i} r_{ij}^{-2}.
    \end{equation}
It turns out that WCN generally performs better than CN at predicting ER \cite{yeh_site-specific_2014, yeh_local_2014, huang_mechanistic_2014}. However, WCN is clearly \emph{not} a local measure. By virtue of the exponent $-2$, the aggregated influence of all amino acids at distance $r$ on WCN is the same, irrespective of $r$. In other words, the factor $1/r^2$ exactly compensates the geometrical increase in the number of residues in each shell, for arbitrarily large radii $r$. Thus, rather than just measuring local packing density, WCN also measures where the focal amino acid falls relative to the overall shape and distribution of amino acids in the protein. For example, in a perfectly spherical protein of uniform density, WCN would be a direct proxy of the distance to the geometric center of the protein.

Moreover, just as was the case with CN, the definition of WCN contains a somewhat arbitrary parameter \cite{yang_protein_2009}, the exponent $-2$. If we chose a more negative exponent, then WCN would be a more local measure of packing density, putting less weight on increasingly distant amino acids. Likewise, if we chose a less negative exponent, WCN would put more weight on more distant amino acids and less weight on amino acids close in. Therefore, just as we had done with CN, we  calculated the correlations between WCN and ER as a function of the exponent $\alpha$ (Figure \ref{fig:cnwcnp}B). We found that the correlations peaked at $\alpha\sim -2.3$, almost exactly the canonical value of $-2$.

\begin{figure}
        \begin{center}
            \includegraphics[width=6in]{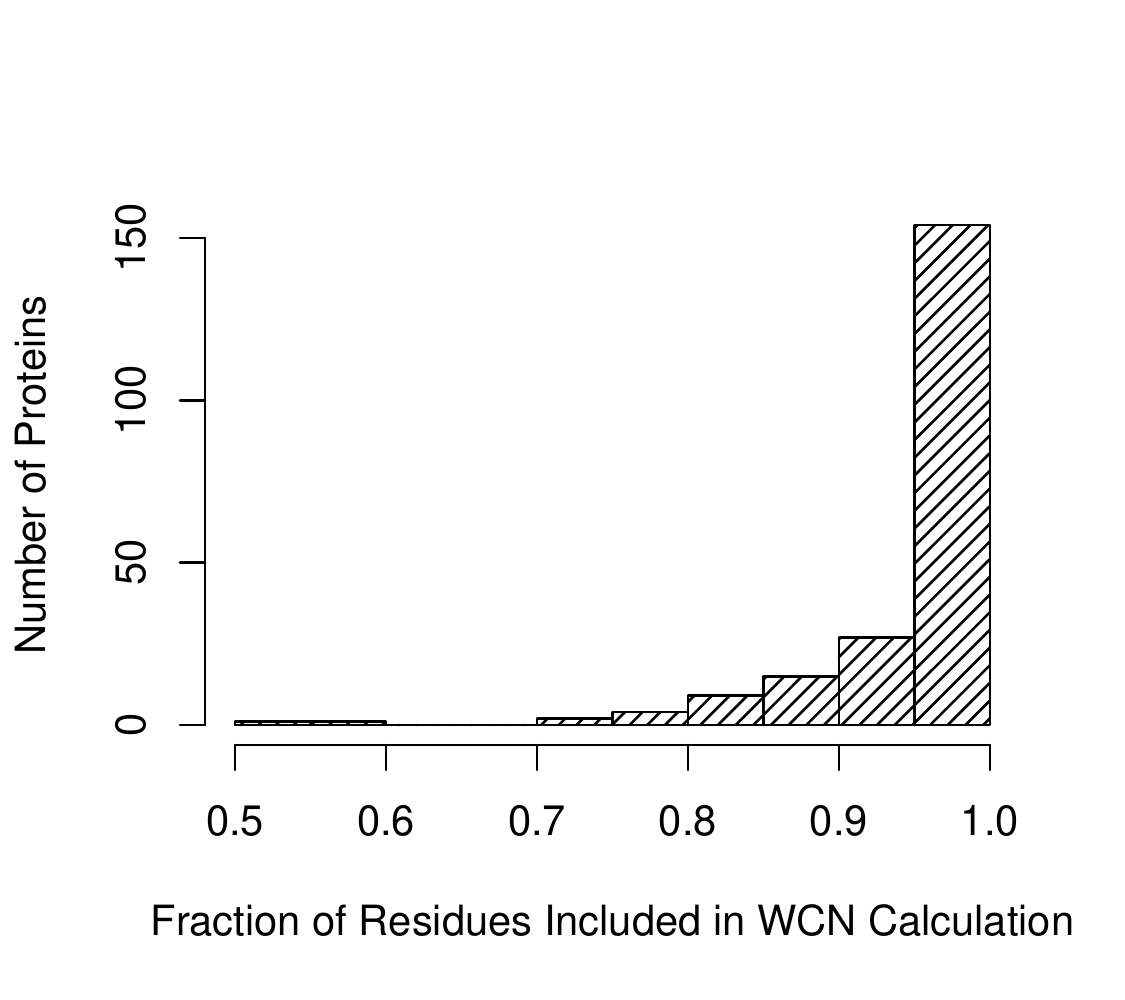}
        \end{center}
        \caption{Fraction of residues that are included in the calculation of $\text{WCN}^\text{cutoff}$ when its correlation with site-specific evolutionary rate is maximized. For the vast majority of proteins, the highest correlation strengths are obtained when over 90\% of the residues in the protein are included in the calculation of $\text{WCN}^\text{cutoff}$.}
        \label{fig:wcn_pwrl_cutoff}
    \end{figure}

Next we asked whether the improved performance of WCN over CN was caused specifically by the power-law weighting of close contacts or by the inclusion of long-range interactions. To do so, we considered a hybrid WCN model with power-law weighting and a hard cutoff beyond which all other residues were excluded. More specifically, we fixed the exponent of the power-law weighting function to $\alpha=-2$ and further included only those sites in the calculation of WCN that were within a specific radius of neighborhood $r_0$ from the site of interest,
    \begin{equation}
        \label{eqn:wcn_pwrl_cutoff}
        \text{WCN}_{i}^\text{cutoff} = \sum^N_{j\neq i} r_{ij}^{-2} H\big(r_0-r_{ij}\big),
    \end{equation}
\noindent with $H$ defined by Eqn. \ref{eqn:heaviside}.
We then varied the cutoff $r_0$ and searched for the value that would result in the strongest correlation of WCN with site-specific evolutionary rates, separately for each individual protein in our dataset. Finally, we measured the mean fraction of neighboring sites that were required in the calculation of $\text{WCN}^\text{cutoff}$  to obtain the highest correlation strengths. We found for the vast majority of proteins that the highest correlation strengths were obtained when over 90\% of the sites in the protein were included in the calculation of $\text{WCN}^\text{cutoff}$ (Figure \ref{fig:wcn_pwrl_cutoff}). This large fraction is a clear indication of the importance of distant sites in their evolution.

In combination, these results demonstrate that both CN and WCN perform best at describing ER variation when they capture a non-negligible contribution of longer-range effects, beyond the immediate neighborhood of each focal site. To disentangle the contributions of local and longer-range effects to sequence evolution, we next proceeded to develop a measure of packing density that by definition depended on only the immediate neighbors around a single amino acid.

\subsection*{Measures of local packing density based on Voronoi tessellations}

\begin{figure}
        \begin{center}
            \includegraphics[width=6in]{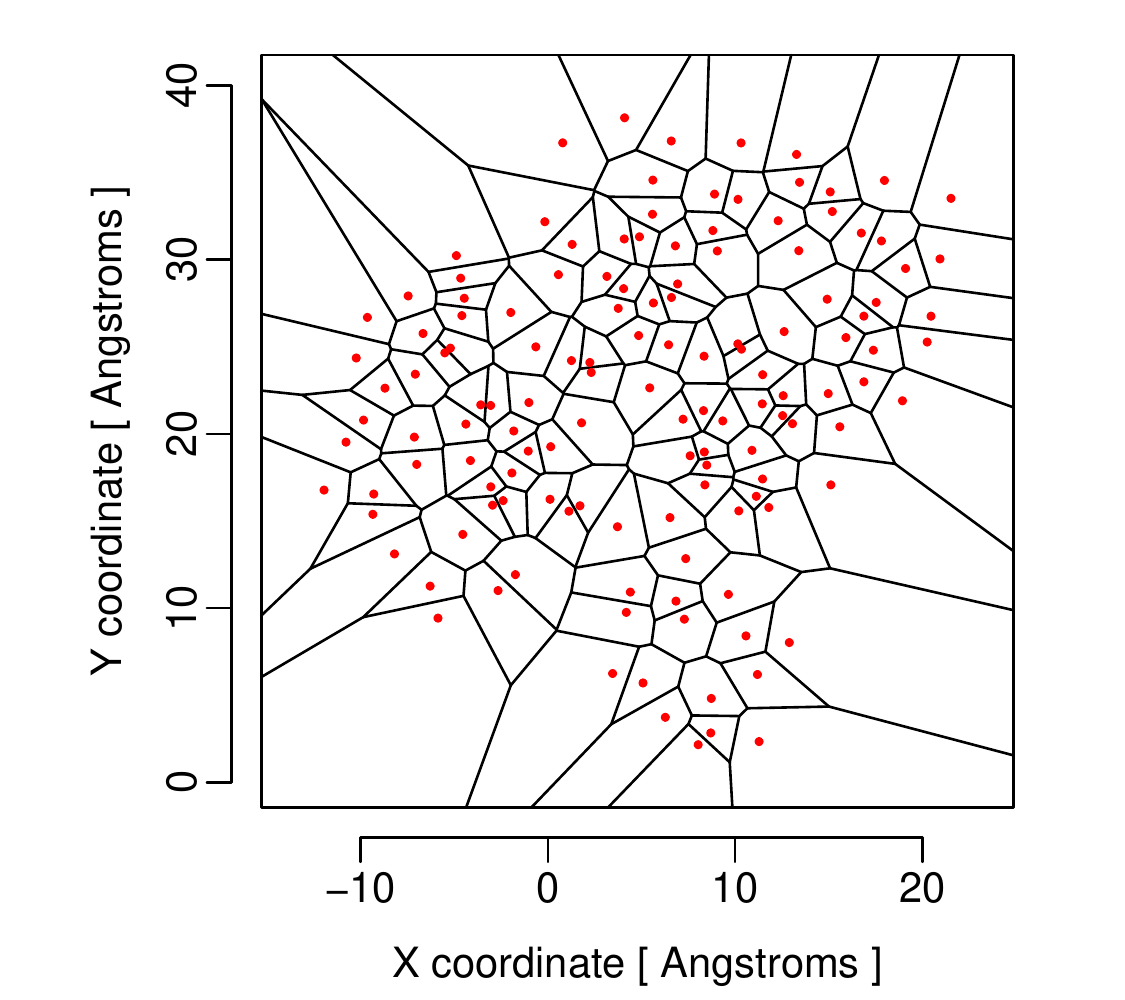}
        \end{center}
        \caption{Example of a Voronoi tessellation in two dimensions. The red dots represent the seed points, and the black lines delineate the Voronoi cells. For protein structures, the tessellation is carried out in three dimensions.}
        \label{fig:voronoi}
    \end{figure}

To describe the local surroundings of a residue inside a protein in an unambiguous manner, we need to partition the space inside a protein into regions that uniquely belong to specific residues. Such structural partitioning has a long history in the analysis of protein structures \cite{richards_interpretation_1974, gerstein_volume_1994}. In particular, the Voronoi tessellation and its dual graph, the Delaunay triangulation, have been used in several studies analyzing the internal structure of proteins and/or developing empirical potentials \cite{zomorodian_geometric_2006, zhou_alpha_2014, xia_identifying_2014}.
The Voronoi tessellation divides the Euclidean space into regions, called {\it cells}, and each belonging to exactly one centroid point (seed), such that the cell corresponding to each centroid point consists of every region in space whose distance is less than or equal to its distance to any other centroid point (Figure~\ref{fig:voronoi}).

When applying a Voronoi tessellation to a protein structure, we usually choose one seed point per residue in the structure. As is the case with CN and WCN, there are multiple possible choices for which point best represents each residue. While many authors simply use the locations of the C$_\alpha$ backbone atoms, we can also use the geometric center of the amino-acid side-chain, the geometric center of the entire amino acid, or any particular heavy atom in the amino-acid backbone or side-chain. We show in Appendix A that we obtain the strongest correlations of Voronoi cell properties with site-specific sequence variability of proteins if we use the geometric centers of the residue side chains as Voronoi seeds. Thus, we used these specific seeds for the remainder of this work.

\begin{figure}
        \begin{center}
            \includegraphics[width=6.5in]{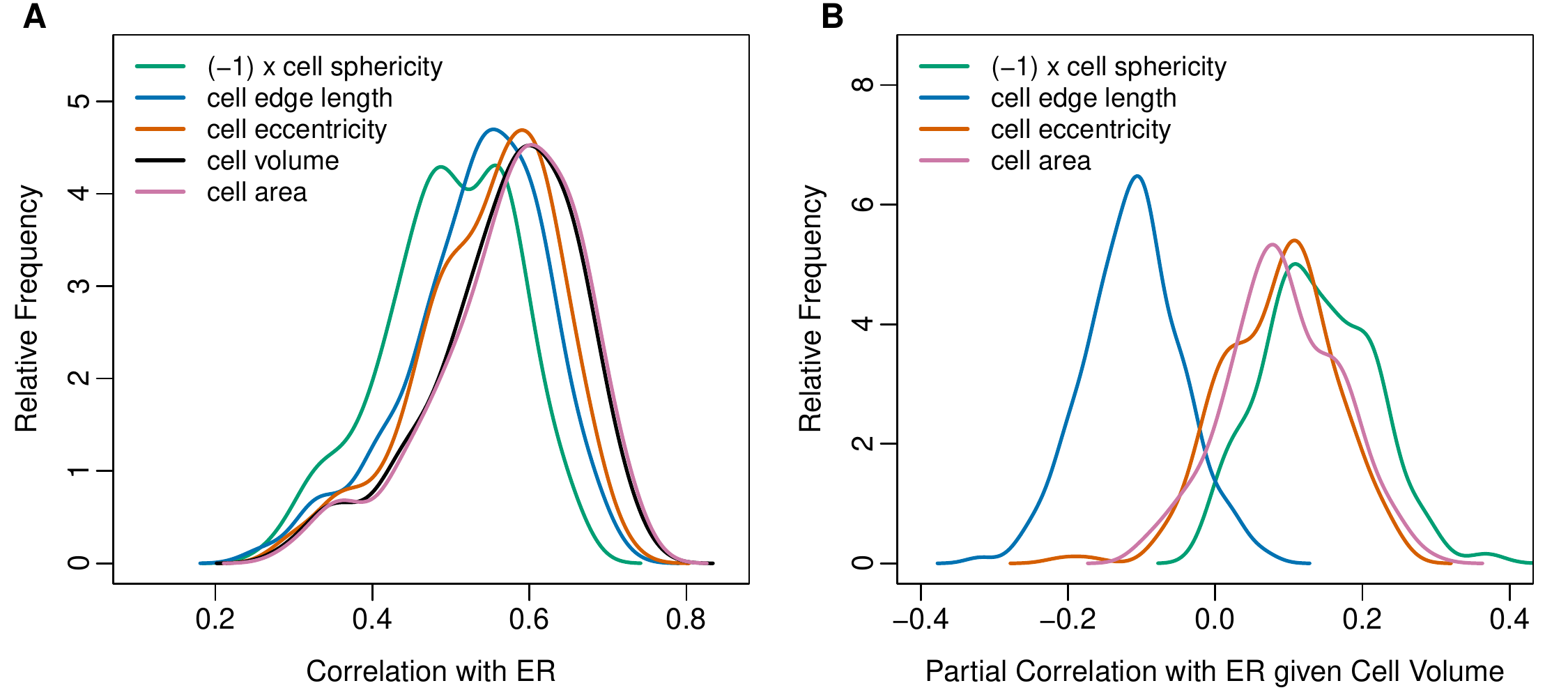}
        \end{center}
        \caption{Correlations and partial correlations of ER with various Voronoi cell properties. (A) Distributions of ER correlations with Voronoi cell properties sphericity, edge length, eccentricity, volume, and area. Note that all cell characteristics correlate positively with ER, except sphericity which correlates negatively with ER. We show here the correlations of ER with $(-1)\times \text{cell sphericity}$ so that the correlation distributions all appear on the same positive scale. The correlations for cell area and cell volume are not significantly different from each other, and are higher than the correlations for all other Voronoi cell properties (Table~\ref{tab:ttest-Voronoi}). (B) Distributions of partial correlations of ER with Voronoi cell properties, controlling for cell volume. All partial correlations are relatively small, with medians of approximately 0.1. Therefore, none of these cell properties provide much independent information about ER once cell volume is accounted for.}
        \label{fig:voronoi_ER_screen}
    \end{figure}

How can the Voronoi tessellation be used to describe the local packing density in a protein? Most importantly, the volume of a Voronoi cell is a direct measure of the amount of space surrounding a particular residue. More densely packed regions of the protein will yield smaller cell volumes than less densely packed regions. However, in addition to cell volume, several other cell properties also measure the local geometric arrangement of residues surrounding the focal one, such as cell edge length, cell area, cell eccentricity, and cell sphericity (see Methods for definitions of these quantities). We asked which of these quantities could be used to predict ER, and we found that they all did (Figure \ref{fig:voronoi_ER_screen}A). Specifically, we found that cell volume and surface area were most strongly correlated with ER, followed by the cell eccentricity, total edge length, and the cell's sphericity (Figure \ref{fig:voronoi_ER_screen}A and Table \ref{tab:ttest-Voronoi}). All of the Voronoi cell characteristics except sphericity correlated positively with ER, while sphericity displayed negative correlations.

    \begin{table}[htbp]
    \caption{\label{tab:ttest-Voronoi}$P$-values of pairwise paired t-tests of the absolute correlation strengths of evolutionary rates (ER) with different Voronoi cell characteristics: volume, area, total edge length, sphericity, eccentricity. Cell area and cell volume perform equally well in predicting ER, while all other cell characteristics perform significantly worse (Fig.~\ref{fig:voronoi_ER_screen}). The $P$-values were corrected for multiple testing using the Hochberg correction \citep{hochberg_sharper_1988}.}
    \bigskip
    \centerline{
    \begin{tabular}{lcccc}
      \hline
                         & ER -- Area        & ER -- Eccentricity & ER -- Edge        & ER -- Sphericity \\ \hline
      ER -- Volume       & $6\times10^{-20}$ & $10^{-26}$         & $2\times10^{-21}$ & $1\times10^{-60}$  \\
      ER -- Sphericity   & $2\times10^{-70}$ & $2\times10^{-43}$  & $4\times10^{-88}$ & --  \\
      ER -- Edge         & $5\times10^{-86}$ & $8\times10^{-15}$  & --                & -- \\
      ER -- Eccentricity & $6\times10^{-37}$ & --                 & --                & --  \\
      \hline
    \end{tabular}}
    \end{table}

In this context, we emphasize that most Voronoi cell characteristics increase with decreasing local packing density. For example, the larger the cell volume or cell surface area, the fewer amino acids are located in the immediate neighborhood of the focal residue. By contrast, increasing sphericity implies increasing packing density. In general, a polyhedron with a higher number of faces corresponds to higher sphericity, the limiting case of which is a perfect sphere with an infinite number of faces. In the context of Voronoi polyhedra in proteins, a higher number of faces of a cell indicates more amino-acid neighbors around the site of interest, and therefore more interactions among them. Consequently, correlations with cell sphericity tend to have the opposite sign from correlations with other Voronoi cell characteristics.

Not unexpectedly, we found that the Voronoi cell characteristics correlated strongly with each other. Therefore, we next used cell volume (measuring inverse local packing density) as the reference and calculated the partial correlations of ER with Voronoi cell properties controlling for cell volume. We found that these partial correlations were generally quite low, with median absolute values of approximately 0.1  (Figure \ref{fig:voronoi_ER_screen}B). Cell edge length and cell sphericity displayed negative partial correlations with ER while cell eccentricity and cell area displayed positive partial correlations.

\subsection*{Influence of longer-range effects on sequence evolution}

\begin{figure}
        \begin{center}
            \includegraphics[width=6.5in]{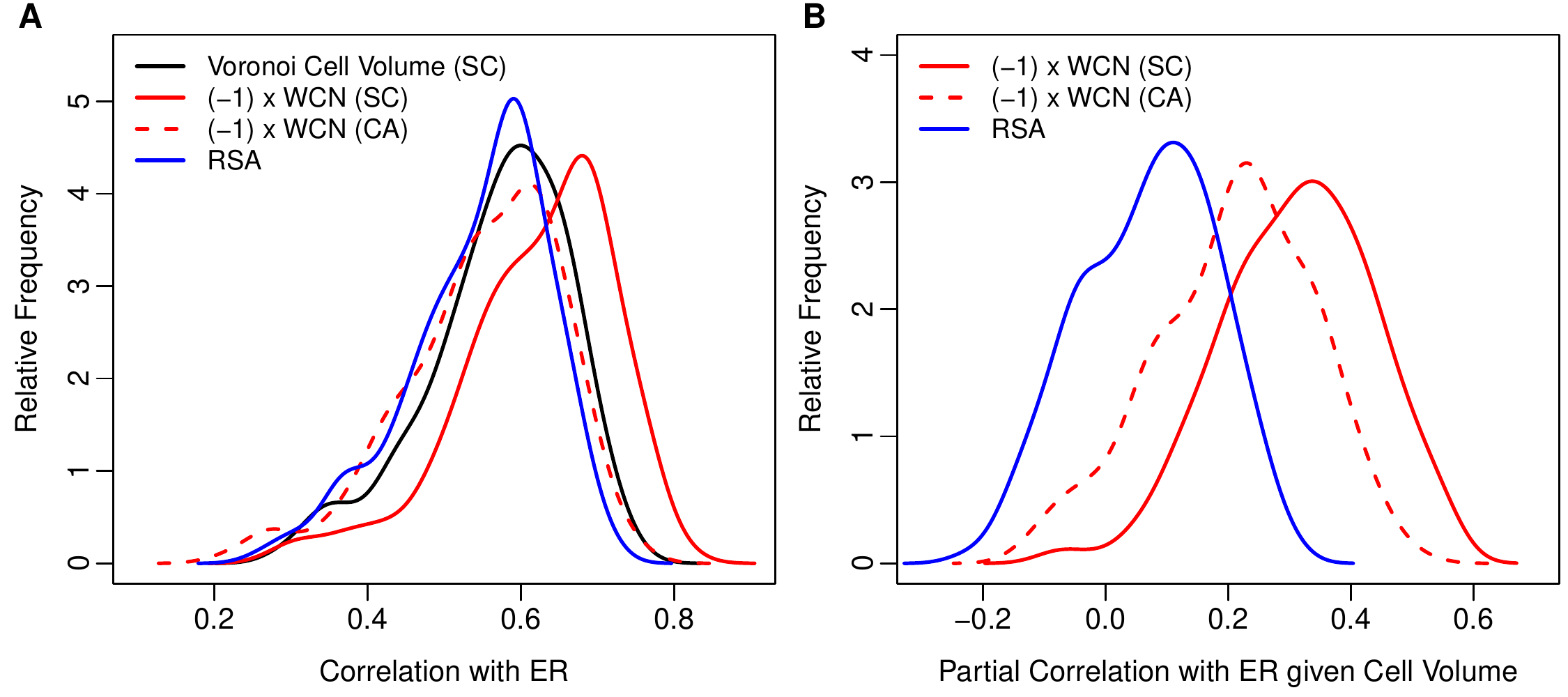}
        \end{center}
        \caption{Correlations and partial correlations of ER with Voronoi cell volume, WCN, and RSA. (A) Distributions of ER correlations with Voronoi cell volume, WCN (using side-chain and C$_\alpha$  coordinates, denoted by SC and CA, respectively), and RSA. Note that the WCN measures correlate negatively with ER. We show here correlations of ER with $(-1)\times \text{WCN}$ so that the correlation distributions all appear on the same positive scale. The correlations of ER with WCN (SC) are significantly higher than the correlations of ER with all other quantities shown (Table~\ref{tab:ttest-best}). (B) Distributions of partial correlations of ER with WCN and RSA, controlling for cell volume. The partial correlation of ER with $-\text{WCN}$ are substantial, with median values of 0.32 (side-chain WCN) and 0.21 (C$_\alpha$ WCN). By contrast, the partial correlations of ER with RSA largely vanish, with a median value of 0.09.} \label{fig:best_predictorER}
    \end{figure}

We next asked how Voronoi cell volume compared as predictor of ER relative to the more commonly studied quantities WCN and RSA. As seen in Figure \ref{fig:best_predictorER}A, Voronoi cell volume, RSA, and C$_{\alpha}$ WCN all displayed comparable correlations with ER, with median (absolute) correlation coefficients of 0.59, 0.57, and 0.56, respectively. By contrast, side-chain WCN out-performed these predictors by a substantial margin, with median (absolute) correlation coefficient of 0.64 (see Table~\ref{tab:ttest-best} for pairwise paired t-tests among all distributions). The superior performance of side-chain WCN on this data set is consistent with previous independent results in the literature \cite{marcos_too_2015}.

    \begin{table}[htbp]
    \caption{\label{tab:ttest-best}$P$-values of pairwise paired t-tests of the absolute correlation strengths of evolutionary rates (ER) with four structural characteristics depicted in Figure \ref{fig:best_predictorER}A: $\text{WCN (SC)}$, $\text{WCN (CA)}$, cell volume, and RSA. The $P$-values were corrected for multiple testing using the Hochberg correction \citep{hochberg_sharper_1988}.}
    \bigskip
    \centerline{
    \begin{tabular}{lccc}
      \hline
         & ER -- WCN(SC) & ER -- WCN(CA)  & ER -- Volume\\ \hline
     ER -- RSA     & $2\times10^{-12}$  & $0.083$ & $6\times10^{-17}$  \\
     ER --  Volume  & $7\times10^{-6}$ & $5\times10^{-4}$ & --  \\
     ER -- WCN(CA) & $5\times10^{-11}$ & -- & -- \\
      \hline
    \end{tabular}}
    \end{table}

Note that both Voronoi cell volume and RSA are purely local quantities, determined entirely by the immediate neighborhood of an amino acid. In fact, they correlate strongly with each other, and measure approximately the same quantity (Appendix A). By contrast, as mentioned previously, WCN takes into account the amino-acid arrangement in the entire protein, and it equally weighs each shell with radius $r$ around the focal residue. Thus, to disentangle local from non-local effects, we calculated partial correlations of ER with WCN while controlling for Voronoi cell volume. In this way, we could quantify the residual influence of WCN---that is, the influence of longer-range effects beyond immediate neighbors---on sequence variability. As a control, we performed the same calculation for RSA instead of WCN.

We found that side-chain WCN displayed substantial residual correlations with ER after controlling for Voronoi cell volume (Figure \ref{fig:best_predictorER}B, Table \ref{tab:explained_var}). The distribution of (absolute) partial correlation coefficients had a median value of 0.32, with an interquartile range of 0.17 (from 0.23 to 0.40). Thus, after controlling for local packing, the longer-range effects appeared to explain approximately $10\%$ of the site-specific sequence evolutionary rates. Similarly, there were significant residual correlations of C$_\alpha$ WCN with ER after controlling for Voronoi cell volume, with a median (absolute) correlation coefficient of $0.21$. By contrast, the residual correlations of RSA with ER were by-and-large negligible. The median was only $0.09$, with an interquartile range of 0.17 (from $-0.02$ to $0.15$).

    \begin{table}[htbp]
    \caption{The percentage of variance of site-specific evolutionary rates explained by Voronoi cell volume (representing the inverse of  site-specific packing density) and Weighted Contact Number controlling for cell volume (representing longer-range effects). In each case, the first, second (median) and the third quartiles of the percentage distribution are reported.  \label{tab:explained_var}}
    \bigskip
    \centerline{
    \begin{tabular}{lccc}
      \hline
      \multirow{2}{*}{Structural Quantity} & \multicolumn{3}{c}{Percentage of ER Variance Explained} \\ \cline{2-4}
                                           & $25\%$ quartile & median & $75\%$ quartile  \\ \hline
      Voronoi Cell Volume        & $27\%$          & $34\%$ & $41\%$  \\
      WCN, controlling for Cell Volume   & $5\%$           & $10\%$ & $16\%$  \\
      \hline
    \end{tabular}}
    \end{table}

In conclusion, either WCN definition (side-chain or C$_\alpha$) provided a significant, independent contribution to explaining ER when purely local effects of packing were controlled for by Voronoi cell volume. By contrast, RSA did not make an independent contribution once Voronoi cell volume was controlled for, as was expected given these quantities' strong correlations with each other.

\subsection*{Sensitivity of results to evolutionary-rate model matrix}

The evolutionary rates we have used throughout this work were calculated based on a Jukes-Cantor-like (JC-like) model, in which all amino-acid substitutions are assumed to be equally likely and no corrections are made for mutation biases and/or the structure of the genetic code \cite{echave_relationship_2014}. More sophisticated evolutionary models, such as the JTT model\cite{jones_rapid_1992}, the WAG model\cite{WhelanGoldman2001}, or the LG model\cite{LeGascuel2008}, attempt to correct for these factors, and generally yield improved model fit compared to simpler models. To what extent these various models affect rate inferences is not well understood. In our analysis here, we used the same rates that had been used in previous analyses \cite{echave_relationship_2014,marcos_too_2015}, so that our results were directly comparable to these previous results. However, the same enzyme data set we analyzed here has also been studied previously with rates calculated under the JTT model\cite{yeh_site-specific_2014,huang_mechanistic_2014,yeh_local_2014}, so we asked to what extent our conclusions differed among these two sets of evolutionary rates.

We found that, first, for each individual protein in the dataset, the evolutionary rates calculated under the JTT model correlated strongly with the evolutionary rates calculated under the JC-like model. The average Spearman correlation coefficient was $0.97$ with the standard deviation of distribution $\sigma=0.02$ over the entire dataset. Second, the results that we report in Table \ref{tab:explained_var} (percent variance explained by Voronoi cell volume and by WCN controlling for Voronoi cell volume) change only by approximately 2 percentage points when JTT rates are used in place of JC-like rates. Similarly, all other analyses also showed only minor variations upon using JTT rates. Therefore, we conclude that our findings are not strongly affected by the specific model with which evolutionary rates are calculated.

\section*{Discussion}

We have shown that the commonly used measures of packing density, contact density (CN) and weighted contact density (WCN), capture non-local effects of amino-acid packing in a protein. To quantify the magnitude of these effects, we have defined purely local measures of packing density via the Voronoi tessellation, and we have shown that Voronoi cell volume encapsulates most of the relevant information about packing contained in various Voronoi cell characteristics. We have then shown that the variation in site-specific evolutionary rates (ER) can be partitioned into two independent components: variation explained by local, site-specific packing density (as measured by Voronoi cell volume) and variation explained by more long-range effects that extend beyond the first coordination shell around each site in a given protein. The longer-range effects play a non-negligible role in protein sequence evolution and explain on average approximately 10\% of the observed variability in ER in the data set studied here. These 10\% are in addition to and independent of the proportion of variance explained by the site-specific packing density, which explains on average another 34\%. By contrast, the relative solvent accessibility (RSA) has negligible explanatory power for ER once site-specific packing is taken into account via Voronoi cell volume.

One limitation of our study is the choice of the data set, 209 monomeric enzymes that have been previously studied by multiple groups. All enzymes in this data set are relatively small, globular, and have a well-defined active site, usually close to the center of the protein \cite{shih_evolutionary_2012}. It is possible that the weighted contact number, with its radial weighting, performs particularly well on such proteins. For example, in the large and elongated influenza protein hemagglutinin, RSA explains over twice the variation in evolutionary rate than WCN does  \cite{meyer_geometric_2015}. Moreover, active sites and protein--protein interfaces generally induce additional evolutionary constraints, and how these constraints interact with local packing density and longer-range steric interactions remains largely unexplored \cite{Deanetal2002, franzosa_structural_2009, meyer_geometric_2015,Abriataetal2015}. Therefore, future work will have to disentangle the contributions of local packing and longer-range effects in a wider set of protein structures, and it will also have to disentangle these contributions from the evolutionary constraints imposed by active sites and protein--protein interfaces.

   The quantities CN and WCN involve freely adjustable parameters in their definitions, and these parameters control to what extent CN and WCN assess the density of both immediate-neighbor and longer-range amino-acid packing. In principle, one could fine-tune these parameters for each individual protein to obtain the strongest correlations between ER and either CN or WCN \cite{yeh_local_2014}. Although the best parameterization of CN and WCN varies from one protein structure to another, we can take the values that maximize the median correlation for our dataset as the best overall parameterization of CN and WCN for our dataset. For both quantities, we find that the optimum parametrization is not entirely local. In particular, for WCN, which is the overall best predictor of ER in this dataset, the optimum exponent is $-2.3$, very close to the canonical value of $-2$ that is commonly used in the definition of WCN  \cite{lin_deriving_2008, yang_protein_2009, huang_mechanistic_2014, yeh_site-specific_2014, yeh_local_2014, marcos_too_2015}. Since an exponent of $-2$ gives nearly equal weight to all shells of any radius around the focal residue, WCN is not really a site-specific measure of packing density. Instead it measures both the local packing density around the residue and the position of the residue relative to the global shape of the entire protein. By contrast, the Voronoi cell volume introduced here as an alternative measure of site-specific packing density depends exclusively on contributions from the nearest neighbors of each site in protein, that is, the first coordination shell around each focal residue.

When searching for both the optimal cutoff for CN and the optimal exponent for WCN, we found a single global optimum in either case (Figure \ref{fig:cnwcnp}). This result was not unexpected. First, the weighting functions (Heaviside step function for CN, power-law for WCN) are monotonically decreasing functions of pairwise distance. Second, even though the pairwise interaction strengths of amino acids in a protein are complex functions of inter-residue distance, virtually unique to each pair of residues in each protein, the overall ``averaged over protein-medium'' behavior of interaction strength vs.\ inter-residue distance decays monotonically with distance \cite{dehouck_effective_2013,xia_correlation_2015,xia_multiscale_2015}. Therefore, the observed behavior in Figure \ref{fig:cnwcnp} merely assert that there is a single optimal parameter value that represents the best relative weighting of local and longer-range interactions. Either increasing or decreasing this parameter value will yield a sub-optimal relative weighting of local and longer-range interactions.

    Several earlier works \cite{yeh_local_2014, huang_mechanistic_2014} have argued that local packing density is the best predictor of ER, while others \cite{franzosa_structural_2009, shahmoradi_predicting_2014} have implicated RSA as the main determinant of sequence evolution, with local packing density playing a more  peripheral role. Importantly, all these works have defined local packing density as CN or WCN. We have shown here that if we define local packing density as the inverse of the Voronoi cell volume, which by definition excludes the effects of longer-range effects beyond the first coordination shell, then it performs comparable to RSA.  Our conclusion is that RSA and local packing density---when measured in a way that excludes longer-range effects---in principle represent the same characteristics of the local environment in a protein, that is, both quantities are proxy measures of the number of neighboring amino acids in the first coordination shell. Additional evolutionary variation is explained by non-local effects and/or the distribution of amino acids at larger distances, and these effects are captured by the most commonly used definition of WCN with inverse $r^2$ weighting.

What is the origin of these non-local effects? There are two potential causes. First, longer-range effects can originate from steric interactions of local clusters of amino acids that extend beyond the first coordination shell. Evidence for such clusters comes from the discovery of protein sectors, regions of proteins with shared co-variation but little co-variation with other regions \cite{Halabietal2009}. Further evidence comes from the observation that correlations between fluctuations of pairs of sites in proteins decay inversely with distances of the sites from each other \cite{dehouck_effective_2013}. Second, evolutionary variation may be influenced by shape and finite-size effects. If proteins have approximately uniform amino-acid density (as our work suggests, see next paragraph), then WCN measures primarily where in the three-dimensional protein structure a given residue is located. Thus, if there are systematic trends of evolutionary variation related to shape and finite size, for example conserved active sites that are located near the center of the protein, then WCN can partially capture these effects and provide further explanatory power for ER.

Other residue--residue interactions besides steric interactions are less likely to cause the long-range effects we have observed here. Covalent bonds and van der Waals' interactions hardly reach beyond the first coordination shells of amino acids, limiting the effective ranges of these interactions to the amino acid's immediate neighborhood $\lesssim10$\AA. And disulfide bonds and ionic interactions are infrequently observed in our sample of enzymatic proteins.

The finding that RSA and Voronoi cell volume are highly correlated was unexpected. According to their definitions, these two quantities measure different aspects of a residue's local environment. In particular, RSA can only vary to the extent to which a residue is actually near the surface. For all completely buried residues $\text{RSA}=0$, with no variability. By contrast, Voronoi cell volume depends on the exact location and number of amino-acid neighbors, and hence can, at least in principle, vary substantially inside the protein. The congruence of RSA and cell volume suggests that cell volume is approximately constant for buried residues and is determined primarily by solvent exposure on the protein surface. These conditions can only be satisfied if the amino-acid density is approximately constant throughout the protein. Thus, the strong correlations we have found between RSA and cell volume imply that amino-acid density is, to first order, uniform across protein structures.

\section*{Appendix A: Average Side-Chain Coordinates as the Best Representation of Protein 3D Structure}
\label{app:best_crd}

The calculations of the quantities CN and WCN, as well as the Voronoi tessellation, all require us to represent each amino acid by a reference point in 3D space. Most commonly, that reference point is chosen to be the C$_\alpha$ atom of the residue. However, this choice is mainly driven by convenience and convention, and \emph{a priori} there is no particular reason to assume that this set of atomic coordinates is the best representation of individual sites in a protein. Indeed, earlier works have already suggested to use the center-of-mass of side chain coordinates to represent residues \cite{soyer_voronoi_2000}. More recently, it was shown that WCN when calculated from side-chain center-of-mass coordinates generally results in significantly better correlations with ER than WCN calculated from $C_\alpha$ atoms does \cite{marcos_too_2015}.

To identify the best reference point in a more principled manner, we here considered seven different choices of atomic coordinates for the calculation of CN, WCN, and Voronoi cells. These include the set of coordinates of all backbone atoms (N, C, C$_\alpha$, O) and the first heavy atom in the amino acid side chains (C$_\beta$). In addition, representative coordinates for each site in the protein can be also calculated by averaging over the coordinates of all heavy atoms in the side chains (referred to as SC). Finally, we calculated a representative coordinate by averaging over all heavy atom coordinates in both the side chain and the backbone of the amino acid (referred to as AA). For all calculations that required C$_\beta$ coordinates, if the side chain $C_\beta$ atom was not been resolved in the PDB file or the amino acid lacked $C_\beta$ (i.e., Glycine), the $C_\alpha$ coordinate for the corresponding amino acid was used instead.

\begin{figure}
        \begin{center}
            \includegraphics[width=6.5in]{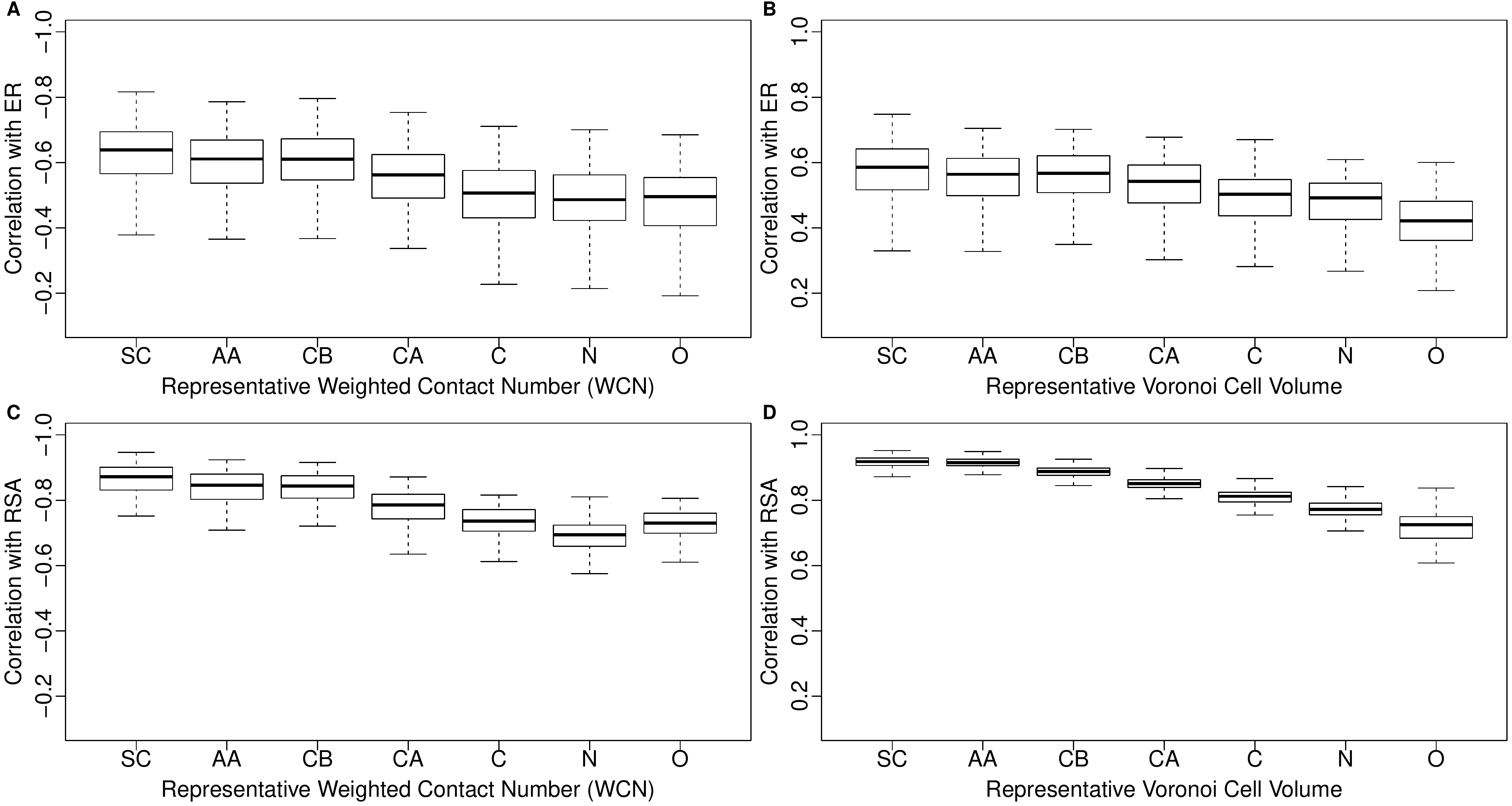}
        \end{center}
        \caption{Side-chain centers provide most informative reference points for both WCN and Voronoi tessellation. (A) Distribution of correlations of WCN with ER, for seven different coordinate sets according to which WCN was calculated: SC, AA, CB, CA, N, C, O. Each coordinate set represents a different way of identifying the reference location of each residue. For SC (Side Chain) and AA (entire Amino Acid), the reference point is given, respectively, by the geometric average coordinates of the Side Chain (SC) atoms and the entire Amino Acid (AA) atoms. The latter include the backbone but exclude any Hydrogen. The coordinate sets CB, CA, N, C, and O use the respective atom in the amino acid as the reference point. (B) As in (A), but using Voronoi Cell Volume instead of WCN. (C) As in (A), but the correlations are calculated with RSA instead of with ER. (D) As in (B), but the correlations are calculated with RSA instead of with ER.}
        \label{fig:best_wcn_vvol}
    \end{figure}

For all seven coordinate choices, we correlated both ER and RSA with both WCN and with Voronoi cell volume for each protein (Figure \ref{fig:best_wcn_vvol}). In all cases, we found a systematic trend of declining correlation strengths as we moved the reference point further away from the geometric center of the side chain. In fact, the two atoms furthest away from the side chain, the nitrogen in the amino-group and the oxygen in the carboxyl group, generally produced the weakest correlations both with ER and with RSA.

Finally, we investigated how Voronoi cell volume related to RSA and WCN. In this analysis, we focused on SC coordinates and considered different exponents for WCN. First, we found that RSA and cell volume were strongly correlated, with a mean correlation coefficient of 0.92 and a standard deviation of $0.02$ (see also Figure \ref{fig:best_wcn_vvol}D). Under the default definition of WCN, with power-law exponent $\alpha=-2$, we found that the mean correlation coefficient between WCN and cell volume was 0.89, with a standard deviation of 0.04. Thus, on average, the correlations between cell volume and RSA were 0.03 higher than the correlations between cell volume and side-chain WCN. We next identified the power-law exponents that resulted in the strongest correlations between WCN and cell volume for each protein. We found a mean exponent of $\alpha=-3.3$ (standard deviation $\sigma=0.3$). Using WCN with the best exponent value for each protein instead of fixed value $\alpha=-2$ for all, the correlation between WCN and cell volume increased to $0.94$ on average (standard deviation $0.01$). These results show that the default WCN definition, with exponent $-2$, is less similar to Voronoi cell volume than a more local definition with an exponent near $-3.3$.

To summarize, our results show that the biophysically relevant reference point for amino-acid position is the side chain, not the commonly used C$_\alpha$ atom. Further, they highlight that RSA properly accounts for the position of the side chain, and behaves nearly identically to side-chain Voronoi cell volume, whereas WCN additional contains information about longer-range interactions not accounted for in the Voronoi cell volume.

\section*{Acknowledgements}

We thank Julian Echave, Benjamin Jack, Austin Meyer, Stephanie Spielman, and Eleisha Jackson for helpful discussions and comments. This work was supported in part by NIH Grant R01 GM088344, DTRA Grant HDTRA1-12-C-0007, and the BEACON Center for the Study of Evolution in Action (NSF Cooperative Agreement DBI-0939454). The Texas Advanced Computing Center at UT Austin provided high-performance computing resources.

\bibliographystyle{unsrt}
\bibliography{main}

\end{document}